\documentclass{p34style}

\def\bx{\mathbf{x}}
\def\bX{\mathbf{X}}
\def\Nev{{\cal N}_{\rm ev}}
\def\meanpt{{\langle p_\bot \rangle_e}}

\newsavebox{\arbx}
\newcommand{\verylongarrow}[1]{
  \savebox{\arbx}(#1,3)[br]{\vector(1,0){#1}}
  \begin{picture}(#1,3)
    \put(0,3){\usebox{\arbx}}
  \end{picture}
  }  

\begin{document}

\title{EXPERIMENTAL CORRELATION ANALYSIS:\\
FOUNDATIONS AND PRACTICE
\thanks{
Proceedings of the 30th International Symposium on Multiparticle
Dynamics, Tihany, Hungary, October 2000, ed.\ T.\ Cs\"org\H{o}
and W.\ Kittel, World Scientific (to be published).}
}

\author{H.C.\ Eggers}

\address{Department of Physics, University of Stellenbosch,
  Stellenbosch, South Africa}

\maketitle

\abstracts{Commonalities and differences in correlation analysis in
  terms of phase space, conditioning and uncorrelatedness are
  discussed. The Poisson process is not generally appropriate as
  reference distribution for normalisation and cumulants, so that
  generalised statistics in terms of arbitrarily defined reference
  processes must be employed. Consideration of the sampling hierarchy
  leads us to a classification of current event-by-event observables.}

\section{Introduction}

In theory, knowledge of the fully differential probability
distribution at all points contains everything there is to know about
a given reaction. While this statement may be true, it is meaningless
in practice because experimental samples are invariably too small to
access even a fraction of the required information; also, the
dimension of phase space is large, creating problems of projection and
visualisation.

In this situation, multiparticle correlations provide meaningful
answers through inclusive sampling: projectiles $A$ and $B$ collide to
yield $N$ final-state particles\footnote{
  We shall pretend that multiparticle final states consist of one
  particle species only.
  }, of which the observer considers only $q$ particles at a time, $A
+ B \longrightarrow p_1 + p_2 + \cdots + p_q + X$, while ignoring the
rest $X$ by means of final-state phase space integration. In the
simplest case $q{=}0$, one considers the multiplicity and the size of
the available phase space,\cite{And88a} advancing to one-particle
distributions ($q{=}1$), two-particle correlations ($q{=}2$) and so
forth.  Every step on the $q$-ladder presents new and increasingly
subtle challenges, both physicswise and in statistical sophistication.
This review hopes to highlight both the simplicity and unity that a
rigorous mathematical approach has taught us.

\section{Phase space and conditioning}
\label{phssp}

Data to be analysed typically consists of a sample $S$ of events $e =
1,2,\ldots,\Nev$, each with a different multiplicity $\hat N^e$ of
tracks, each of which has momentum and other measured quantum numbers
contained in a vector $X_i^e, i = 1,\ldots \hat N^e$. For $q$-th order
correlation measurement, the sample is completely described by the set
of \textit{per-event counter densities}, one for each
event,\footnote{Notations $\ \hat{}\ $ and $e$ shall be omitted and used
  interchangeably as is opportune.}
\begin{equation}
\label{cnb}
\hat\rho_q(x_1,\ldots,x_q) 
= \sum_{\stackrel{\scriptstyle i_1 \neq i_2 \neq \cdots}
                 {\scriptstyle \neq i_q = 1}}^{\hat N}
f(X^e_{i_1},X^e_{i_2},\ldots, X^e_{i_q}) \;
\delta(x_1 - X^e_{i_1}) \cdots \delta(x_q - X^e_{i_q}) \,,
\end{equation}
which registers a value $f_1(x_1,\ldots,x_q)$ whenever the coordinates
$(X^e_{i_1},\ldots,X^e_{i_q})$ of an ordered data $q$-tuple exactly
match the observation points $(x_1,\ldots,x_q)$.  Some simple examples
for $\hat\rho_q$ and $f$ are the all-charge rapidity counter
$\hat\rho_1(y) = \sum_{i=1}^{\hat N} \delta(y - y_i)$ with $f\equiv
1$, and, using a 4-dimensional $\mathbf{X}^e_i \equiv
(y_i,\phi_i,p_{\bot\,i},c_i)$ with charges $c_i = \pm 1$, a
second-order positives-only counter $\hat\rho_2(\bx_1^+,\bx_2^+) =
\sum_{i\neq j} \delta_{c_i,1}\delta_{c_j,1} \; \delta(\bx_1 - \bX^e_i)
\; \delta(\bx_2 - \bX^e_j)$ or a charge-charge counter
$\hat\rho_2(\bx_1,\bx_2) = \sum_{i\neq j} c_i c_j \; \delta(\bx_1 -
\bX^e_i) \; \delta(\bx_2 - \bX^e_j)$.

\medskip

Essentially all correlation measures can be represented as
combinations of the two kinds of averages over
$\hat\rho_q(x_1,\ldots,x_q)$ shown in Figure~1:
\begin{enumerate}
\item ``spatial'' averaging, in the form of \textbf{phase space
    integrals} $\int_\Omega dx_1 \, dx_2 \, \ldots dx_q$ of the
  arguments $x$ over a selected region of phase space $\Omega$, and
\item \textbf{sample averaging} $\langle \ \rangle_S = (1/\Nev)\sum_e$
  over a sample $S$.
\end{enumerate}
\begin{figure}[h]
  \thicklines \setlength{\unitlength}{1pt}
  \begin{picture}(250,80)
    \put(50,0){
      \begin{picture}(200,80)
        \put(000,60){\framebox(70,20){$\hat\rho_q(x_1,\ldots,x_q)$}}
        \put(150,60){\framebox(70,20){$\hat N^{[q]}(\Omega)$}}
        \put(000,00){\framebox(70,20){$\rho_q(x_1,\ldots,x_q)$}}
        \put(150,00){\framebox(70,20){$\left\langle N^{[q]}(\Omega)
                                       \right\rangle_S$}}
        
        \put( 35,60){\vector(0,-1){40}}
        \put(185,60){\vector(0,-1){40}}
        \put( 40,40){$\langle\ \rangle_S$}
        \put(190,40){$\langle\ \rangle_S$}
        
        \put(70,10){\vector(1,0){80}}
        \put(80,15){$\int_\Omega dx_1 \cdots dx_q$} 
        \put(70,70){\vector(1,0){80}}
        \put(80,75){$\int_\Omega dx_1 \cdots dx_q$} 
        
      \end{picture}
      }       
  \end{picture}
  \caption{Phase space and event averaging of the per-event counter
    $\hat\rho_q$ yields, respectively, the per-event factorial moment
    $\hat N^{[q]} = \hat N(\hat N{-}1)\cdots(\hat N{-}q{+}1)$, the
    correlation density $\rho_q(x_1,\ldots,x_q)$ and the sample
    factorial moment $\langle N^{[q]}\rangle$ for a particular choice
    of $\Omega$ and $S$.}
\end{figure}
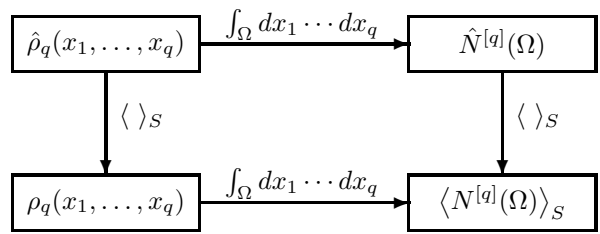

\noindent
Along with the choice of variable $x$ and weight function $f$, the
choice of phase space $\Omega$ and sample $S$ over which to average is
motivated by the physics under investigation\footnote{Experimental
  limitations such as acceptance and limited statistics also play a
  role.}. Choice of $\Omega$ concerns particle selection; choice of
$S$, called \textit{conditioning}, concerns event selection.

\begin{figure}[t]
  \centerline{
    \parbox[t]{7pc}{\epsfxsize=7pc  \epsfbox{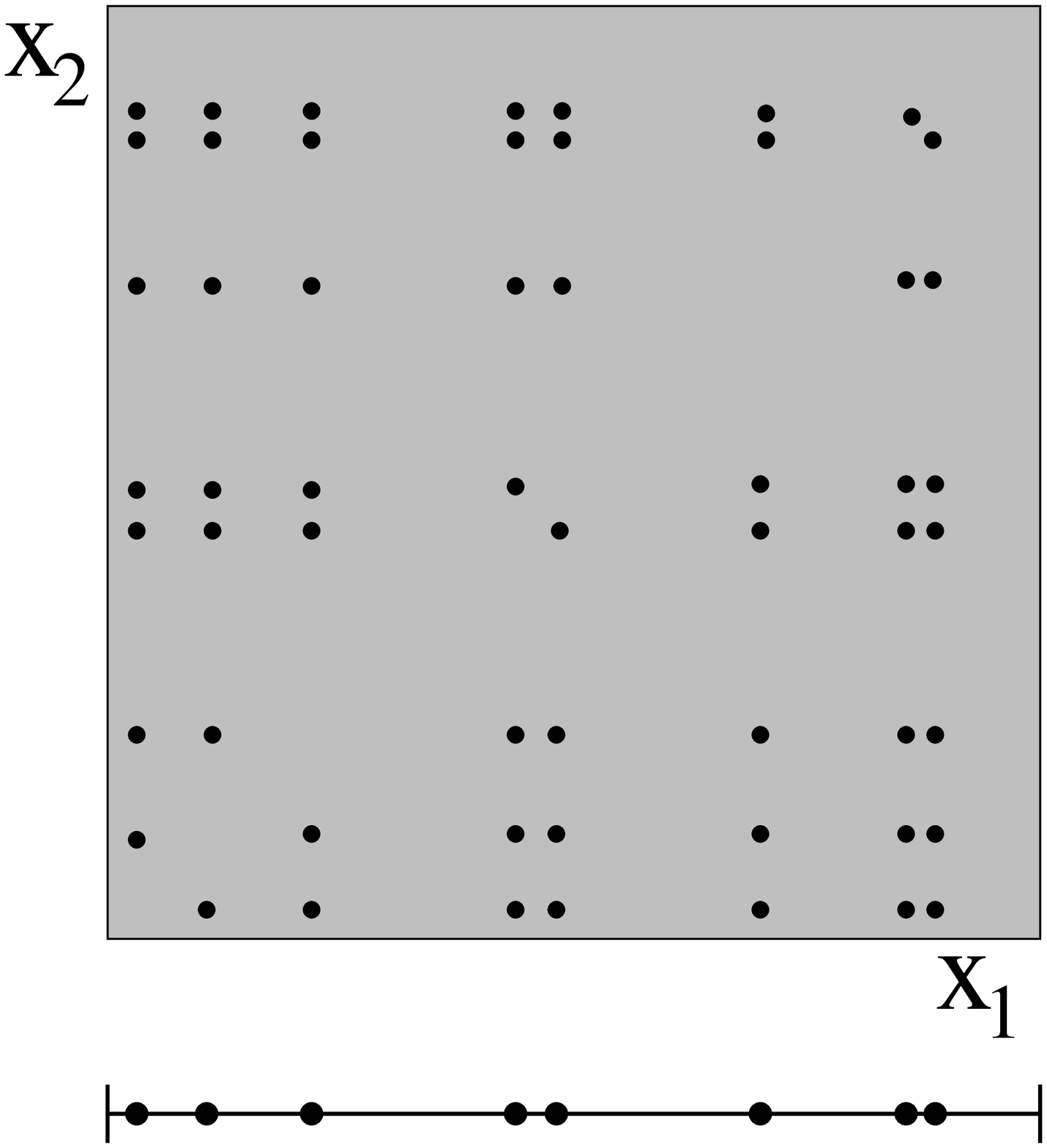}} 
    \parbox[t]{7pc}{\epsfxsize=7pc  \epsfbox{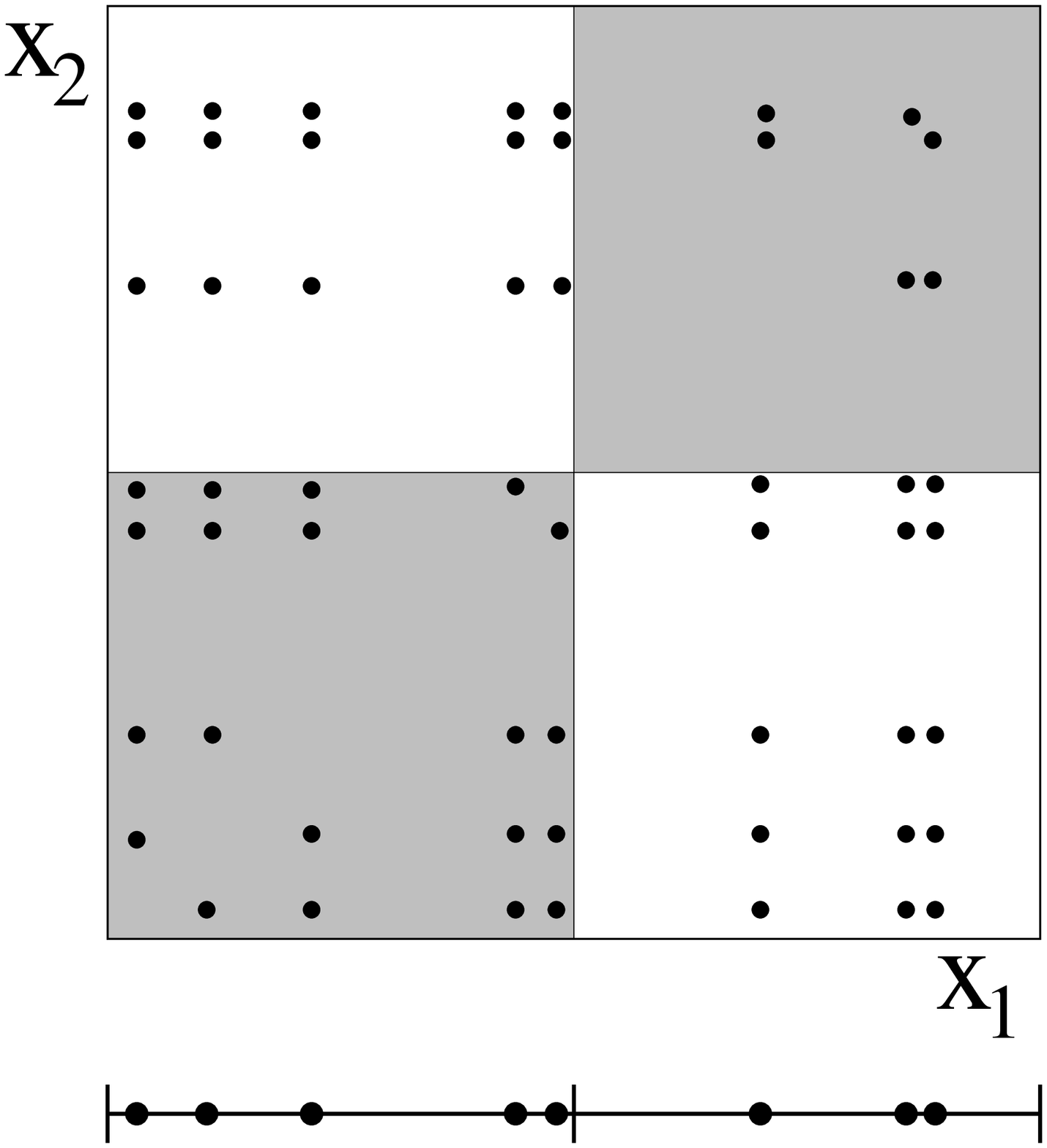}} 
    \parbox[t]{7pc}{\epsfxsize=7pc  \epsfbox{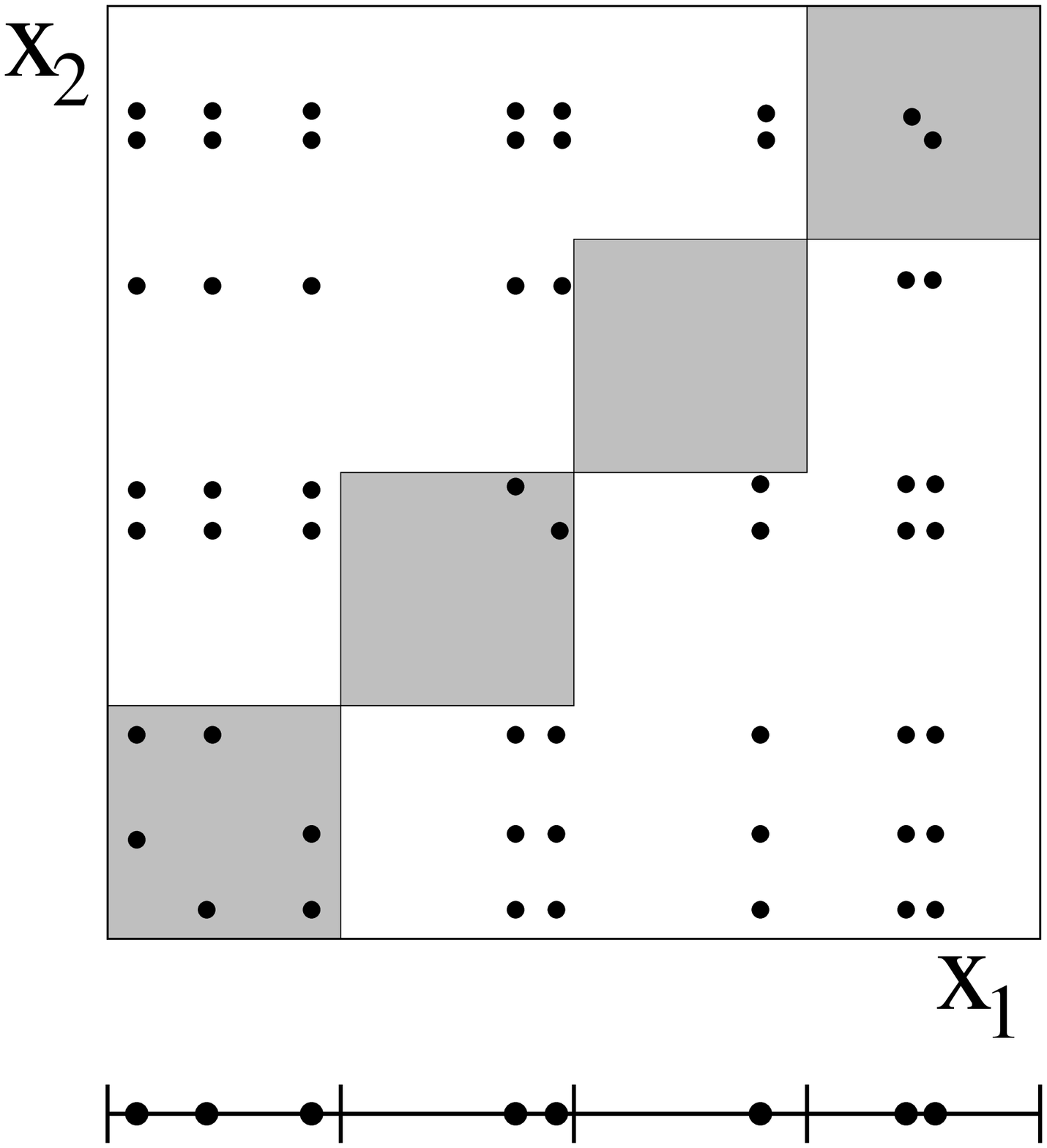}} 
    \parbox[t]{7pc}{\epsfxsize=7pc  \epsfbox{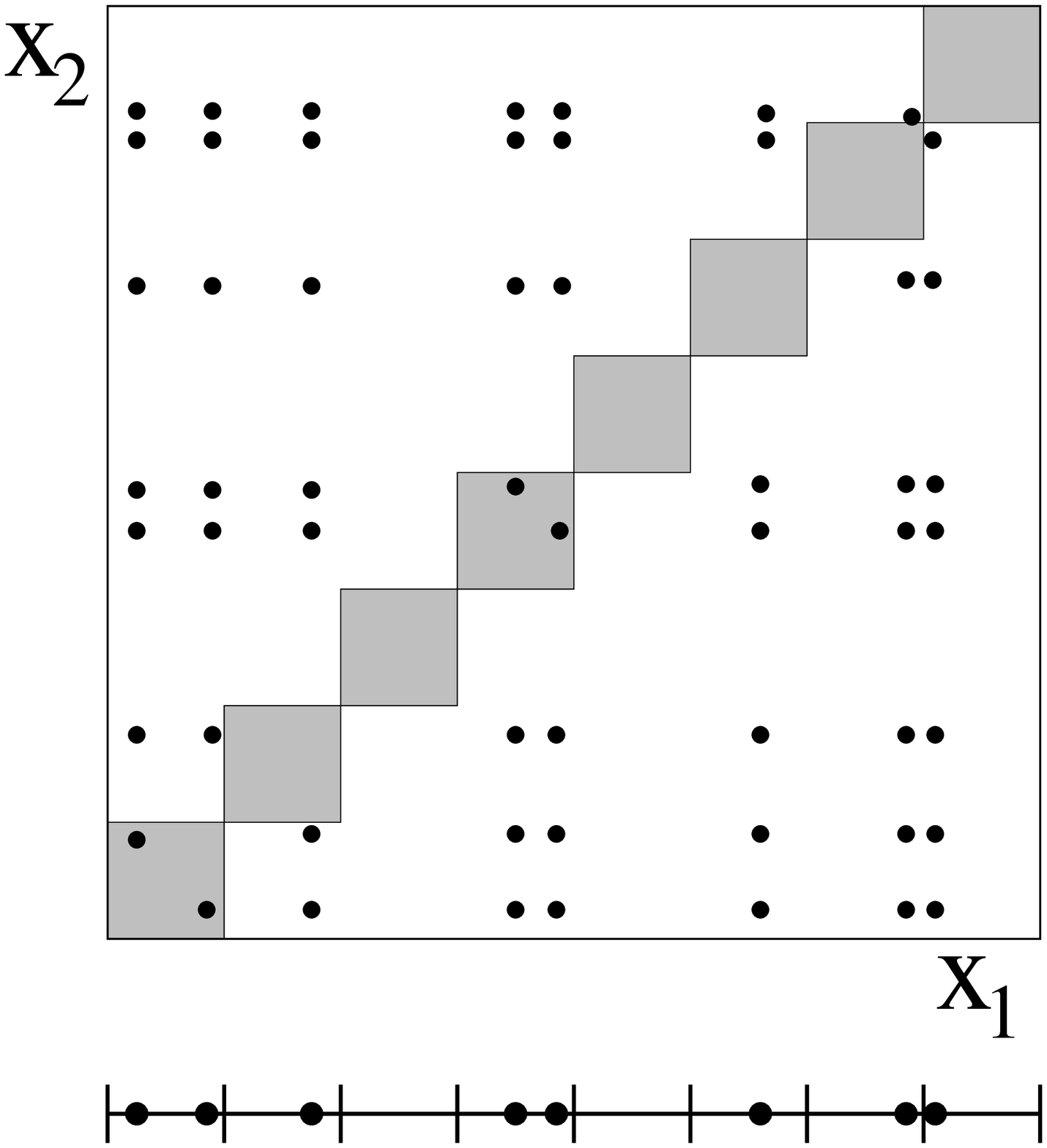}}
    }
  \par\vspace*{-18pt}
  \caption{Phase space integration domains $\Omega$ (shaded areas) for various
    stages of box intermittency analysis. Each point represents a
    particle 2-tuple (pair) of the event whose tracks are represented
    as dots on the line below the phase space plot. Also shown is the
    corresponding increasingly fine binning of the event.}
\end{figure}

\noindent
We focus first on $\Omega$. In Figure~2, we show, for the old
box-moment intermittency analysis, the appropriate $\Omega$'s as
shaded areas, for various resolutions $L/2^j$, superimposed on an
example counter $\hat\rho_2(x_1,x_2)$, each of whose dots represents a
pair drawn from the $\hat N = 8$ tracks shown on the line below the
plot itself.  Further examples of $\Omega$ for $\hat\rho_2(y_1,y_2)$
for fixed $y_1$ and the differential correlation
integral~\cite{Egg93a} over the same counter are shown in Figs.\ 3a
and 3b respectively; greyscales represent the $\Omega$'s for
successive data points.  The correlation density
$\rho_q(x_1,\ldots,x_q) = \langle \hat \rho_q \rangle$, which for
$q=2$ is conveniently represented as a two-dimensional contour plot,
is shown schematically in Figs.\ 3c and 3d for rapidity $y$, together
with $\Omega$'s for forward-backward and rapidity gap measurements.
\par\vspace*{-2mm}\par
\begin{figure}[h]
  \centerline{
    \parbox[t]{7pc}{\epsfxsize=7pc  \epsfbox{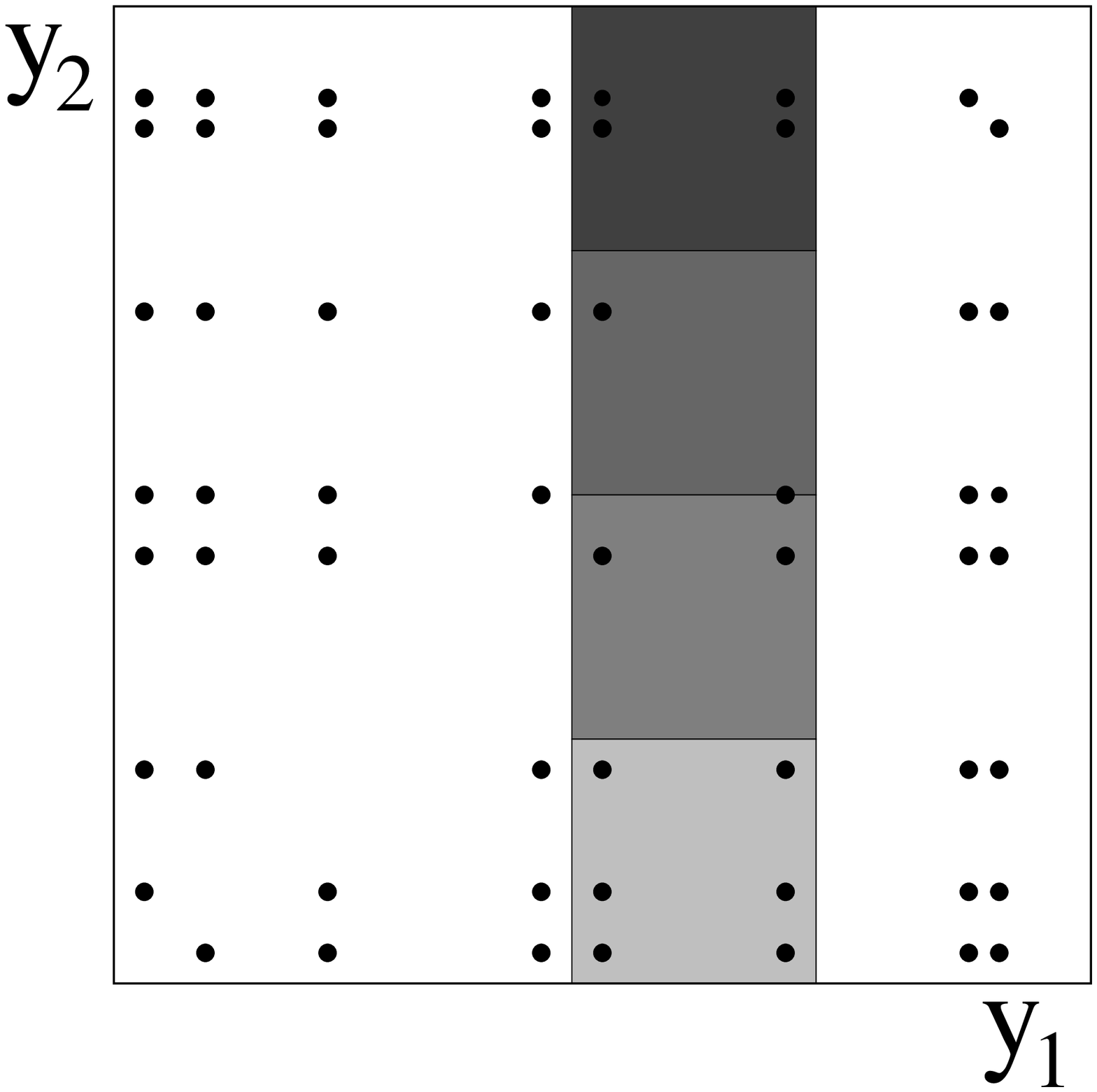}}
    \parbox[t]{7pc}{\epsfxsize=7pc  \epsfbox{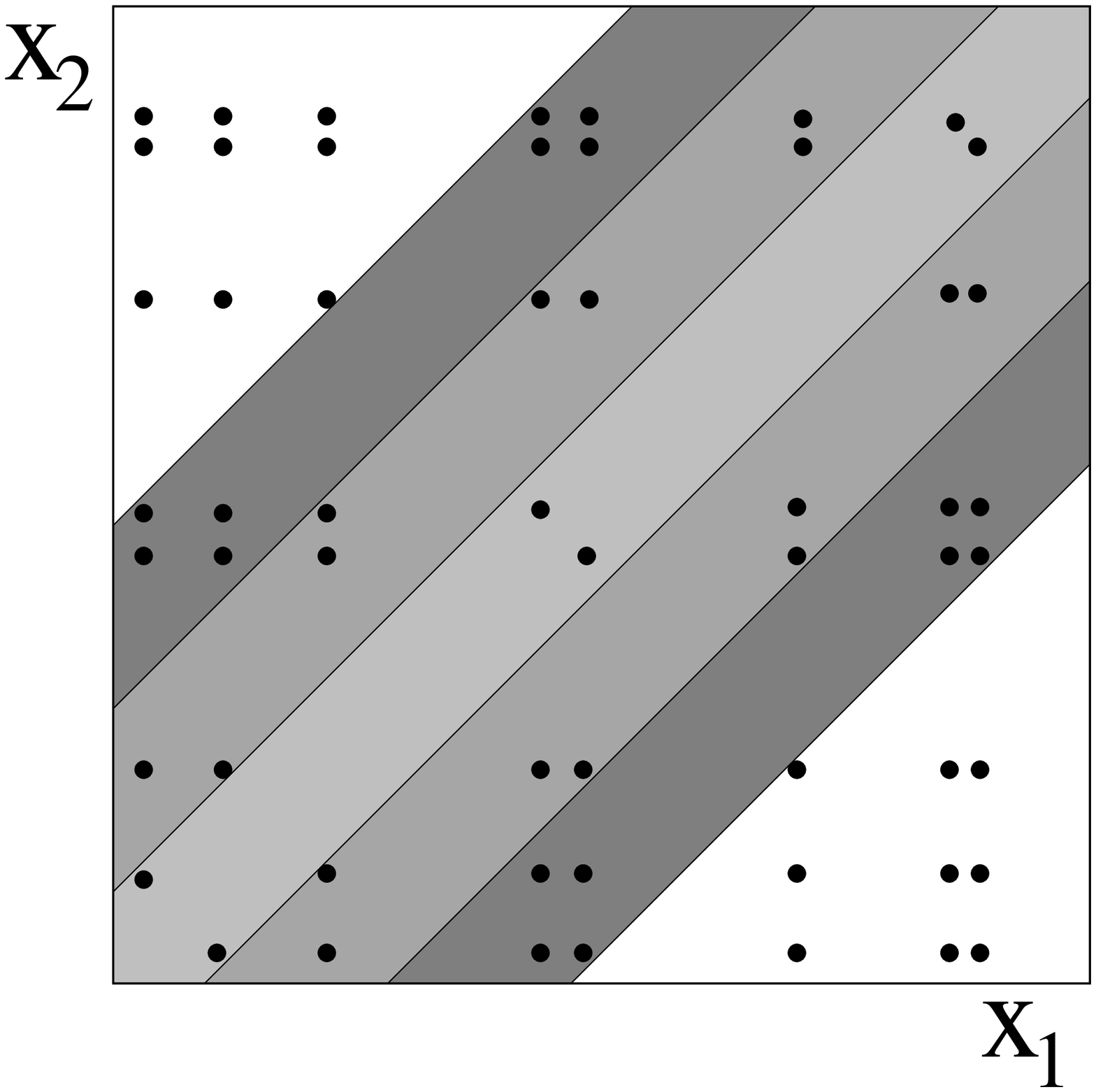}}
    \parbox[t]{7pc}{\epsfxsize=7pc  \epsfbox{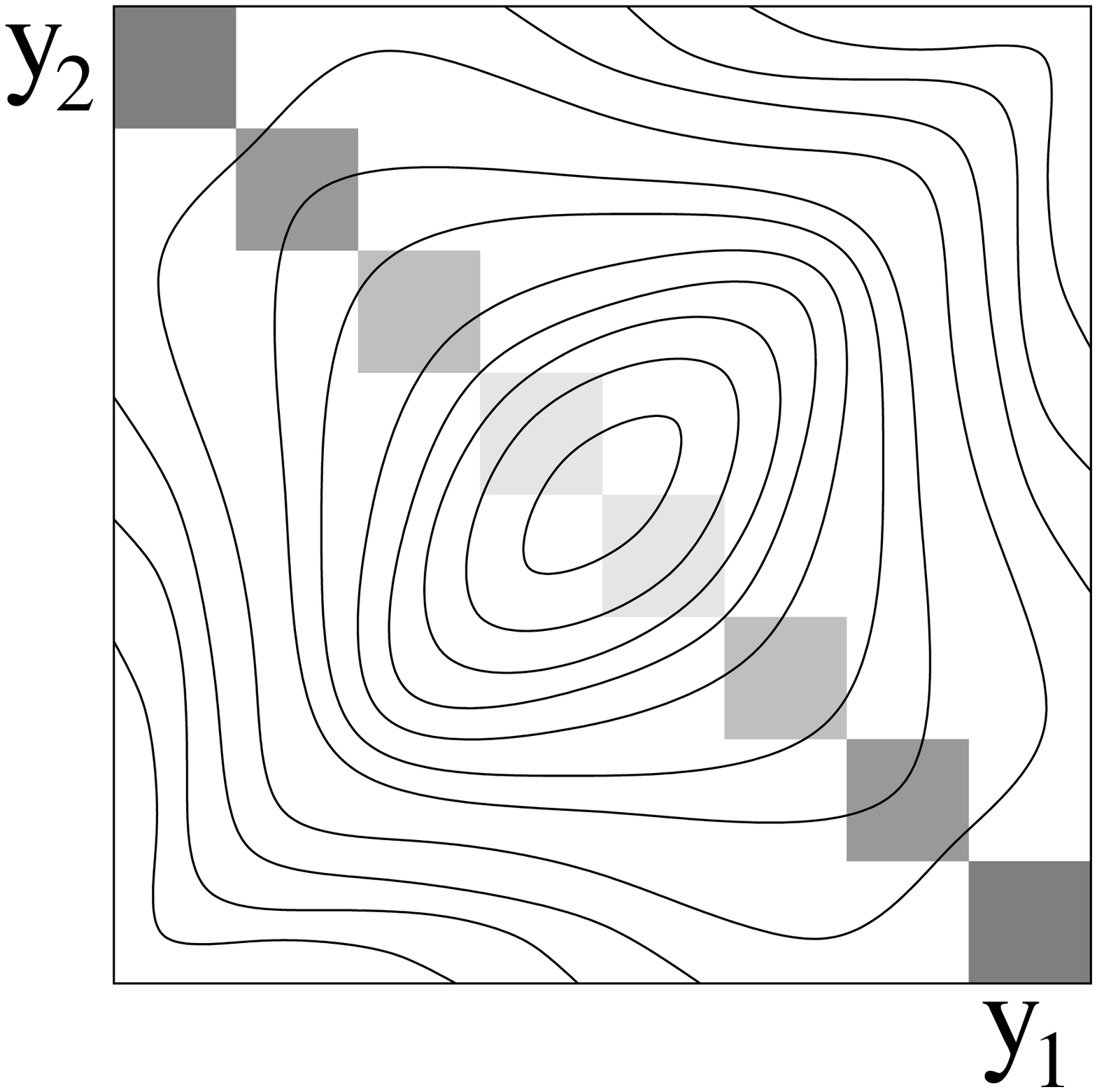}}
    \parbox[t]{7pc}{\epsfxsize=7pc  \epsfbox{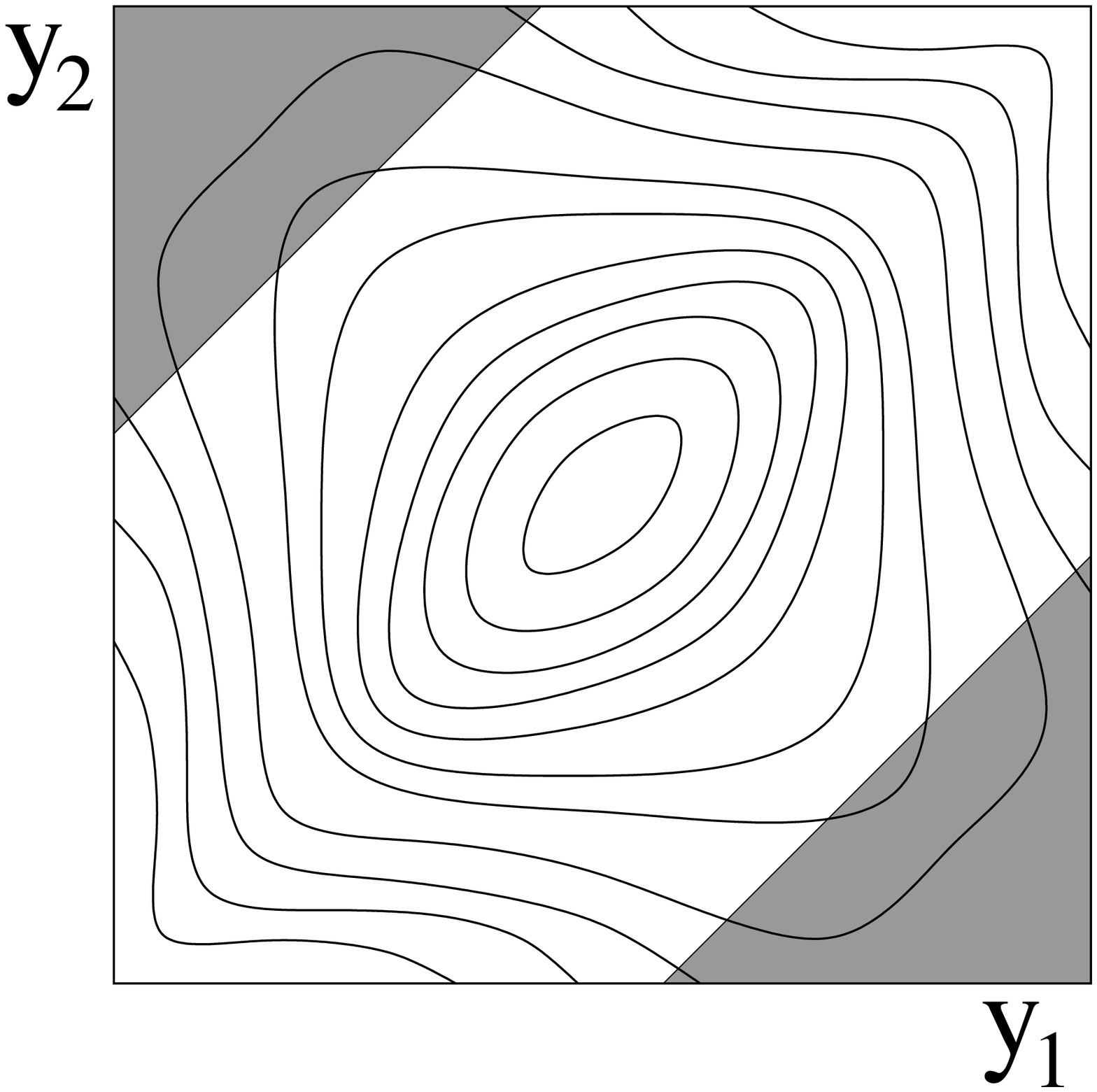}}
    }
  \vspace*{-7mm}\par
  \hspace*{11mm}
  \parbox[b]{29.5mm}{\scriptsize (a)}
  \parbox[b]{29.5mm}{\scriptsize (b)}
  \parbox[b]{29.5mm}{\scriptsize (c)}
  \parbox[b]{10mm}{\scriptsize (d)}\\
  \vspace*{-5mm}
  \caption{Examples of phase space integrals over $\Omega$. 
    From left to right: 
    (a) box correlations keeping rapidity $y_1$ fixed; 
    (b) correlation integral with linear binning in $|x_1 - x_2|$; 
    (c) forward-backward correlations for fixed bin widths but 
        varying bin-bin separation; 
    (d) rapidity gap correlations. 
    }
\end{figure}

\noindent
There is freedom of choice in sample averaging also. ``Conditioning''
is the choosing of a subsample of events from the available full
sample according to some criterion $I_C$ based on attribute $\hat C$,
such as an interval $I_C \equiv \{C\in [C_{min},C_{max}] \}$.
Examples of conditioning are overall multiplicity ($\hat C = \hat
N(\Omega_{\rm tot})$), tagging ($\hat C$ = some quantum number),
central collision selections, and kaon-to-pion ratio.

The influence of conditioning appears to be obvious and quite limited,
in that different (sub)samples will clearly yield different results,
and so little attention has been paid to it. This is deceptive,
however: conditioning, or rather its absence, directly enters the
foundations of correlation analysis as a hidden assumption and is
therefore of central importance.

\section{Uncorrelatedness}

Experimental correlation analysis needs to be both measurable and
meaningful.  It is measurable only as far as trivial contributions to
the correlated signal can be successfully eliminated; it is meaningful
only to the extent that implicit assumptions are understood and
fulfilled. Elimination of trivialities is accomplished in part by
cumulants and appropriate normalisation, while consideration of an
assumption underlying uncorrelatedness, the absence of correlation,
will lead us to the ``reference distribution''.

\subsection{Cumulants}

Moments $\hat N^q$ and factorial moments $\hat N^{[q]}$, whether
per-event or sample-averaged, contain correlations of all orders $\leq
q$ and so are very bad indicators of correlation of order $q$ itself.
These trivial lower-order contributions are removed by the use of
cumulants, which are thus generally preferable.\footnote{
  There are, to my mind, only two justifications for use of moments:
  the first lies in possible scaling behavior, $N^q(\Omega) \simeq
  \Omega^{\zeta_q}$ with constant $\zeta_q$; the second is that
  moments can be defined on a per-event basis, which cumulants
  generally cannot.}
Important for our purposes are two well-known cumulant properties:
\begin{itemize}
\item \textbf{Behaviour under statistical independence}: A
  differential cumulant (e.g.\ $C_2(x_1,x_2) = \rho_2(x_1,x_2) -
  \rho_1(x_1)\rho_1(x_2)$) becomes zero whenever any one of its $q$
  arguments is statistically independent of (uncorrelated with respect
  to) the others, and the integrated factorial cumulant (e.g.\ 
  $\langle N(N{-}1) \rangle - \langle N \rangle^2$) is zero when the
  distribution in $\Omega$ is poissonian: \setlength{\jot}{9pt}
  \begin{eqnarray}
    \label{cma}
    C_q(x_1,\ldots,x_q) && 
    \stackrel{\mbox{{\ \scriptsize statistical independence\ }}}
    {\verylongarrow{100}}\; 0 \,,
    \\
    \label{cmb}
    \int_\Omega C_q(x_1,\ldots,x_q) \, dx_1 \cdots dx_q  &&
    \stackrel{\mbox{{\ \scriptsize Poissonian in $\Omega$\ }}}
    {\verylongarrow{80}}\; 0 \,.
  \end{eqnarray}
  
\item \textbf{Additivity}: If two distributions $P_f$ and $P_h$ are
  independent, the cumulant of the convolved distribution is the
  arithmetic sum of the individual cumulants, 
  \begin{equation}\label{cni}
    C_q^{(f*h)} = C_q^{(f)} + C_q^{(h)} \,.
  \end{equation}
\end{itemize}
There is a problem, however: While eq.~(\ref{cmb}) is true and seems
simple enough, it assumes implicitly that the sample $S$ (i) is
inclusive, and (ii) has an overall poissonian multiplicity
distribution $P_{\hat N} = e^{-\nu} \nu^{\hat N}/\hat N!$. A sample
for which these assumptions are not fulfilled (i.e.\ almost every
conditioned sample!) will hence not yield zero cumulants, even when it
is otherwise considered ``uncorrelated'' by the naive user.  Indeed,
it has long been a source of consternation that for fixed-multiplicity
samples, $P_{\hat N} = \delta_{\hat N, N_{\rm fix}}$, the cumulant
does not integrate to zero, $\int dx_1\, dx_2\, C_2 = \int dx_1\, dx_2
(\rho_2 - \rho_1^2) = N(N{-}1) - N^2 = -N$; similarly, the
nonpoissonian overall multiplicity of UA1 data is known to be the
source of nonzero $C_2(\Omega_{\rm tot})_{\rm UA1} \simeq 0.4$.

For the stronger condition (\ref{cma}) to be fulfilled, the
uncorrelated sample needs to be a realisation of the \textit{Poisson
  process}, for which both $P_{\hat N}$ is poissonian and all possible
subdivisions of phase space are both mutually uncorrelated and
poissonian in themselves.

\subsection{Normalisation}

Normalisation is designed to get rid of the trivial dependence on the
overall multiplicity (leaving nontrivial multiplicity-driven effects)
and the shape of the one-particle distribution. Both trivial
multiplicity dependence and shape are eliminated by the ``vertical''
or differential normalisation $\rho_1(x_1) \cdots \rho_1(x_q)$, which
is commonly held to be appropriate because for uncorrelated samples
$\rho_q \to \rho_1 \cdots \rho_1$ and so
\begin{equation}
  \label{nmb}
  r_q(x_1,\ldots,x_q) = 
  {\rho_q(x_1,\ldots,x_q) \over \rho_1(x_1) \cdots \rho_1(x_q)}
  \stackrel{\mbox{{\ \scriptsize statistical independence\ }}}
  {\verylongarrow{100}}\; 1\,.
\end{equation}
Again, however, this normalisation turns out to be incorrect for any
conditioned sample; again, the implicit assumption being made in
writing $\rho_q \to \rho_1^q$ is that the uncorrelated case is the
Poisson process, which is clearly impossible once conditioning changes
$P_{\hat N}$. Most experimentalists are intuitively aware of this; for
example, no central collision sample is ever normalised with
poissonian $P_{\hat N}$ but strictly by one constructed with the same
experimental non-poissonian $P_{\hat N}$ governing the 
numerator.\footnote{This takes care, however, only of the overall
nonpoissonian character, but not of internal correlation issues.}

\subsection{The reference distribution}

We have seen that textbook definitions of moments, cumulants and
normalisations mathematically equate uncorrelatedness with the Poisson
process. We also know that the majority of samples are not compared to
the Poisson process at all; on the contrary, all sorts of
nonpoissonian cases are considered trivial or, in our parlance,
definitive of ``uncorrelatedness''.

Moreover, different physics questions addressed to the same data
sample may consider different parts of its correlation structure to be
trivial and nontrivial: some questions need to eliminate kinematical
constraints, some do not; some theorists want to eliminate the effect
of resonances in order to isolate the ``quantum statistics'', while
others need to eliminate quantum statistics to look for dynamical
effects.  One approach wants to get rid of jet effects, another needs
them explicitly.  Different jet-finding algorithms will yield
differently conditioned data samples. And so on.

Clearly, in practice there is not one exclusively valid definition of
the trivial, uncorrelated sample, but as many different ones as there
are interesting experimental questions to ask. Explicit consideration
and definition of the appropriate reference distribution, that
distribution considered by the physics at hand to be trivial, is
therefore not a luxury but the very foundation of making any
meaningful statement on correlation itself.

Once this is accepted, it follows that all measurement quantities
(cumulants, normalisation, \ldots) must also be tailored to the
appropriate reference distribution.  For cumulants, the tailor's job
turns out to be simple: based on discrete Edgeworth expansions, it can
be shown rigorously~\cite{Lip96a} that any process $f$ has cumulants
with respect to any reference process $h$ of $C_q(f {\rm with\ 
  respect\ to\ } h) = C_q(f) - C_q(h)$.  These ``generalised
cumulants'' reduce to the normal ones when $h$ is the Poisson process,
and can be viewed as the deconvolution of $f$ with respect to $h$,
just as eq.~(\ref{cni}) represents the convolution of $f$ with $h$.
For experimental cumulant definitions, $h$ would be defined by the
desired reference distribution, and the tailormade cumulant would be
\begin{equation}
  \label{rfb}
  C_q^{\rm nontrivial\ effect} = C_q^{\rm data} - C_q^{\rm reference}\,.
\end{equation}
With regard to normalisation, it is conceptually important that the
appropriate random variable in $q$-th order phase space is a $q$-tuple
$(X^e_{i_1},X^e_{i_2},\ldots, X^e_{i_q})$ rather than $q$ individual
particles $X^e_i$. The distinction lies exactly in the fact that a
$q$-tuple is easily conceived of as being drawn from a correlated
reference sample, while $q$ individual particles tend to be viewed as
(poisson) uncorrelated. The normalisation (\ref{nmb}) is hence to be
replaced by
\begin{equation}
  \label{rfc}
  r_q(x_1,\ldots,x_q)^{\rm nontrivial\ effect} = 
  {\rho_q(x_1,\ldots,x_q)^{\rm data} \over 
   \rho_q(x_1,\ldots,x_q)^{\rm reference}} \,,
\end{equation}
which by construction tends to unity when the data and reference
moments are the same.  Examples for $\rho_q^{\rm reference}$ are the
usual $\rho_1(x_1)^{\rm data}\cdots \rho_1(x_q)^{\rm data}$ for the
Poisson case, and ${N^{[q]}\over N^q} \rho_1(x_1|N)^{\rm data}\cdots
\rho_1(x_q|N)^{\rm data}$ for the multinomial, the appropriate
reference distribution for fixed-multiplicity samples.  More
generally, when Monte Carlos are used to simulate what is considered
trivial or uninteresting for a particular data set and physics
question, the appropriate reference moment is $\rho_q^{\rm reference}
= \rho_q(x_1,\ldots,x_q)^{\rm MC,\ trivial\ effects}$. This
corresponds exactly to common practice.

Although there is no mathematical theorem for this, the natural
normalisation for cumulants would seem to be the same as that for
moments, so
\begin{equation}
  \label{rfd}
  K_q(x_1,\ldots,x_q)^{\rm nontrivial\ effect} = 
  {C_q^{\rm data} - C_q^{\rm reference}   \over 
   \rho_q^{\rm reference}} \,.
\end{equation}
Obviously, the identical phase space integral $\int_\Omega$ needs to
be applied to both numerator and denominator whenever phase space
averaging is applied.

Two applications will illustrate the points made: the extraction of
inter-$W^+ W^-$ Bose-Einstein correlations \cite{wwexpts} currently
much under discussion~\cite{Dew01a} and the practice of ``double
normalisation''.  In the former case, one regards Bose-Einstein
correlations originating separately within the $W^+$ and $W^-$ jets,
measured in semileptonic (sl) decays, to be trivial; hence $C_2^{\rm
  ref} \equiv C_2^{\rm BE\ in\ sl({+}) jet} + C_2^{\rm BE\ in\ sl({-})
  jet}$, and so the nontrivial, inter-$WW$ correlations in the fully
hadronic $q\bar qq\bar q$ sample would according to (\ref{rfb}) be
\begin{equation}
  \label{rfe}
  C_2^{{\rm inter-}WW} =
  C_2^{{\rm BE\ pairs\ in}\ q\bar qq\bar q\ {\rm sample}}
- C_2^{\rm BE\ in\ sl({+}) jet}
- C_2^{\rm BE\ in\ sl({-}) jet} \,.
\end{equation}
Second, doubly normalised cumulants, generically of the form
\begin{equation}
  \label{rff}
  {
    \left( {\rho_2^{\rm data} \bigl/ \rho_1^{\rm data} \rho_1^{\rm data} }
    \right)
    \over
    \left(
      {\rho_2^{\rm MC} \bigl/ \rho_1^{\rm MC} \rho_1^{\rm MC}}
    \right) 
    }
  \; - \; 1 \,,
\end{equation}
have often been applied heuristically, precisely with the intention of
ridding the data of trivial but correlated reference distributions.
Now, from (\ref{rfd}) we get a nontrivial cumulant of
\begin{equation}
  \label{rfg}
  {C_2^{\rm data} - C_2^{\rm MC} \over \rho_2^{\rm MC}}
  =
  {     \rho_2^{\rm data} - \rho_2^{\rm MC} 
      - \rho_1^{\rm MC}  \rho_1^{\rm MC} 
      + \rho_1^{\rm data}\rho_1^{\rm data}
    \over 
    \rho_2^{\rm MC}
  }\,,
\end{equation}
which can be reduced to the form (\ref{rff}) if and when
$\rho_1(x)^{\rm data} \equiv \rho_1(x)^{\rm MC}$ for all $x$.
Conventionally used double ratios are thus in agreement with our
derivations to the extent that this condition is satisfied. For
higher-order cumulants $q\geq 3$, however, double ratios will always
differ from the ``exact'' cumulant subtraction method.

We mention in passing that the ``statistical fluctuations'' of
Ref.~\cite{Vol99a} are nothing but its reference distribution, (which,
as always, does not define itself but must be defined), while the
$\sigma^2_{\rm stat}$ is the cumulant which the authors subtract out,
in agreement with (\ref{rfb}).

\section{Event-by-event analysis and the sampling hierarchy}

Rather than attempting an overly condensed summary of experimental
results and issues in general,\footnote{The interested reader should
  consult existing reviews such as Refs.~\cite{Dew96a}.} we now
concentrate on so-called event-by-event (or ``EbE'') analysis which
has gained increasing prominence in recent years. Our limited goal is
to show that, as far as the statistics are concerned, approaches and
results called EbE in the literature fall into three distinct classes,
namely true event-by-event observables, sample statistics, and
conditioned statistics. See also Ref.~\cite{Tra00a} for another
classification scheme.

To understand the significance and interrelationship between these
classes, we describe briefly an example of a sampling hierarchy. At
the basis of the hierarchy is the set of \textit{single track
  measurements}, grouped into \textit{events} making up the inclusive
data \textit{sample}, from which the relevant \textit{sample
  statistics} can be extracted. Many samples make up a \textit{sample
  of samples} or metasample, yielding in turn \textit{sampling
  statistics}. Further grouping of metasamples and metametasamples can
clearly be continued ad infinitum, at least conceptually, hopefully
asymptotically approaching what a scientist might accept as
``statistical truth''.

In Fig.~4, this hierarchy and its related correlation structure is
shown.\footnote{
There are, of course, other hierarchies besides the one shown,
including multivariate forms and those where event averaging precedes
phase space averaging.}
Starting in Box 1 from per-event counters, the desired attribute $\hat
C^e, e = 1,\ldots,\Nev$ is measured per event. Direct event averaging
(Boxes $2\to 4$) over $\hat C$ is convenient but also immediately
loses most information. One therefore additionally constructs the
\textit{frequency distribution}, the relative frequency of events for
which observable $\hat C$ takes on a particular value,
\begin{equation}
  \label{rlfr}
  P(\hat C) = P_{\hat C} \equiv {\Nev(\hat C) \over 
         \sum_{\hat C} \Nev(\hat C)}\,.
\end{equation}
This opens up (Box 3) the world of \textit{sample statistics}, the
simplest of which is $P(\hat C)$ itself. Much more information can
thereafter be gleaned from higher-order sample statistics $\sum_{\hat
  C} P_{\hat C} h(\hat C)$ with various functions $h$.  The case $h(\hat C)
= \hat C$ leads us back ($3\to 4$) to the sample mean.

Similarly, direct averaging of sample statistics ($4\to 6$) is
complemented by constructing the \textit{sampling distribution} ${\cal
  P}(\langle C \rangle)$, the relative frequency of samples with a
particular value for attribute $\langle C \rangle$, which then forms
the basis for sampling statistics (Box 5).

Conditioning enters the picture when a subsample or set of subsamples
is selected by some criterion $I_C$. Each resulting \textit{conditioned
  sample distribution} $P(C\,|\, I_C)$ in Box 7, with the appropriate
subsample $S_{I_C} \subset S$, then forms the starting point for a
hierarchy of its own, with a corresponding set of conditioned
statistics and averages.
\\

\noindent
Popular EbE schemes can now be seen as falling into three distinct
classes:
\begin{itemize}
\item[I] \textbf{True per-event observables:} These look directly at
  \textbf{one event only}, considering, for example, statistics of
  that event's internal structure such as per-event factorial
  moments~\cite{Bia86a} $\hat N^{[q]}$, the wavelet-transformed
  density~\cite{Dre00a} and per-event slope parameters. Wavelet
  transforms are of particular interest, because using the wavelet as
  weight function $f$ in eq.(\ref{cnb}) goes well beyond the simple
  step functions or prefactors mostly in use.

  Averages within this class are necessarily restricted to phase space
  (``spatial'', ``horizontal'') averages; this includes any sums over
  particles. Any attempted differential or ``vertical'' normalisation
  will necessarily rely on averages over a conditioned subsample having
  the same property as the event under consideration, but external to
  the event itself. Needless to say, such mixed cases must be handled
  with particular care.
  
  Cumulants, generally reliant on event averages, are not easily
  defined on a per-event basis. Again, relative to some sample, this
  can be done,~\cite{Egg93a} but only with special care.

\item[II] \textbf{Sample statistics:} Quantities in
  Ref.~\cite{Vol99a,Gaz92a,NA49_99a} and similar work are typical
  sample statistics; for example, the observable
  \begin{equation}
    \label{rlfs}
    \sigma^2_{\meanpt}
    \equiv
    \sum_{\meanpt} P_{\meanpt} \meanpt^2 
    - \biggl( \sum_{\meanpt} P_{\meanpt} \meanpt
    \biggl)^2
  \end{equation}
  is the second cumulant of the sample distribution in $\meanpt$, the
  mean transverse momentum of tracks in event $e$. 
  
  Clearly, such quantitites are not truly event-by-event but
  sample-based; the misnomer is probably based on the fact that $P_C$
  is an event ratio. Sample statistics as such are nothing new in the
  sense that multiplicity distributions, scatter plots of
  events~\cite{Wu93a} and generally observables based on relative
  frequencies of events have been in the experimentalist's toolbox for
  a long time. New in Ref.~\cite{Gaz92a} and similar ones is the fact
  that these either go beyond first-order sample statistics or have
  invented novel attributes $\hat C$. Neither their ancestry nor their
  misclassification detracts from the interest and relevance of these
  quantities.

\item[III] \textbf{Conditioned statistics:} This class can be termed
  EbE in that the process of conditioning selects or discards events
  according to event properties with, again, $\hat C$ a fixed
  multiplicity, a trigger based on a single track\cite{MiniMax}, a
  $q$-tuple (e.g.\ invariant mass), the whole event (total transverse
  energy) etc. After selection, however, conditioned measurements can
  vary as much in character as unconditioned ones, ranging from true
  Class I EbE measurements to (conditioned) sample statistics.  As in
  Class II, the novelty lies chiefly in exploiting higher-order
  sample statistics.

\end{itemize}
Nothing forbids the use of observables higher up in the sampling
hierarchy; on the contrary, the sampling distribution in Box 5 forms
the basis for any estimate of the dispersion of sample statistics
(``Were I to repeat my experiment many times, how often would my
currently measured results be matched?''). Examples are
Ref.~\cite{Lip91a}, where sampling statistics of factorial moments
were calculated in order to estimate the reliability of single-sample
moments, and standard variances of means, which are just the
$q{=}2$ cumulants of the sampling distribution.

\section{Conclusions}

Very briefly: The first part of this review intended to show that most
correlation measurements can be understood as various phase space
integrations and event averages over per-event counters $\hat\rho_q$.
Through selection of $\Omega$ and $S_{I_C}$, one zooms in on various
aspects of the overall reaction.

After discovering the Poisson process lurking as an unwarranted
assumption in the normal definitions of cumulants and normalisation,
we found solace in the existence of generalised cumulants and
normalisations which appear successfully to allow for and accommodate
any desirable reference distribution.

The current flurry of event-by-event analysis turns out, on closer
inspection, to consist of three different classes, namely true
event-by-event quantities, sample statistics, and conditioned sample
statistics.  For clarity in this and most questions of experimental
statistics, the appropriate sampling hierarchy is an indispensable
navigation instrument.

\section*{Acknowledgments}
Most of the ideas presented here originated in joint research and
discussions with friends and colleagues B.\ Buschbeck and P.\ 
Lipa.  This work was funded in part by the South African National
Research Foundation.




\clearpage


\begin{figure}[h]
\begin{center}
  \setlength{\unitlength}{1pt}
  \begin{picture}(350,410)

    \put(130,373){\framebox(90,24){
        \vbox{\textbf{Inclusive }\\ \textbf{data sample} $S$}}}
    \put(130,378){\thicklines\vector(-1,0){40}}
    \put(130,392){\thicklines\vector(-1,0){40}}

    \put(000,360){\framebox(12,12){1}}
    \put(000,360){\framebox(90,50){\vbox{\mbox{ }\\[-10pt]
                      {\small per-event}\\[-4pt]
                      {\small counters}\\
                      $\hat\rho_q(x_1,\ldots,x_q)$\\ $\forall\; q,x$  }}}
    \thicklines
    \put( 45,360){\vector(0,-1){40}}
    \put( 05,330){\shortstack[b]{\small phase\\ 
                                 \small space\\ \small averaging}}
    \put( 50,340){\shortstack[b]{$\int_\Omega$}}

    \put(000,260){\framebox(12,12){2}}
    \put(000,260){\framebox(90,60){\vbox{
                       {\small per-event global}\\[-3pt]
                       {\small or local}\\[-3pt]
                       {\small observable}\\[3pt]
                       e.g.\ $\hat C = \sum_i X_i^e$, \\
                       $\hat N = \int_\Omega \hat\rho_q $  }    }}
    \thicklines
    \put( 45,260){\vector( 0,-1){80}}
    \put( 05,210){\shortstack[b]{\small averaging\\ 
                                 \small over\\ \small events}}
    \put( 50,215){${\displaystyle{1\over \Nev}} \sum_e$}

    \put( 90,266){\vector( 2,-1){48}}
    \put(110,260){\small event frequency distribution}

    \put(000,140){\framebox(12,12){4}}
    \put(000,140){\framebox(90,40){\vbox{{\small sample mean}\\
                          $\langle C \rangle$ }            }}
    \thicklines
    \put(45,140){\vector(0,-1){80}}
    \put(05,090){\shortstack[b]{\small averaging\\ 
                                \small over\\ \small samples}}
    \put(48,100){${\displaystyle{1\over N_S}} \sum_S$}
    \put( 90,146){\vector( 2,-1){48}}
    \put(110,140){\small sample frequency distribution}

    \put(000,020){\framebox(12,12){6}}
    \put(000,020){\framebox(90,40){\vbox{{\small sampling mean}\\ 
                   $\langle\langle C \rangle\rangle$ }     }}

    \thicklines
    \put(45,19.5){\vector(0,-1){20}}
    \put(25,09){\small etc.}
    \put(50,09){\small etc.}

    \put(90,30){\vector(2,-1){20}}
    \put(115,10){etc.}

    \put(000,-012){\framebox(90,12){\small \textbf{Statistical Truth}}}


    \put(130,200){\framebox(12,12){3}}
    \put(130,200){\framebox(90,40){\vbox{{\small sample distribution}\\
                     $P(\hat C)$,\\ \hspace*{15pt}\small sample statistics} }}

    \put(140,199.5){\vector(-2,-1){47}}
    \put(110,174){$\sum_C P(C)\, C $}

    \put(220,220){\vector(1,0){44}}
    \put(224,226){\shortstack[b]{\small conditio-\\ ning}}
    \put(224,190){\shortstack[b]{\small with\\ criterion\\ $I_C$}}

    \put(130,80){\framebox(12,12){5}}
    \put(130,80){\framebox(90,40){\vbox{{\small sampling distribution}\\
                                 ${\cal P}(\langle C \rangle)$,\\ 
                                 \hspace*{15pt}\small sampling statistics} }}
    \thicklines
    \put(140,079.5){\vector(-2,-1){47}}
    \put(110,054){$\sum_{\langle C\rangle}
                   {\cal P}(\langle C \rangle)\, \langle C\rangle$}


    \put(264,200){\framebox(12,12){7}}
    \put(264,200){\framebox(86,40){
        \vbox{
          {\small conditioned}\\
          {\small sample distribution}\\
          $P(\hat C\, |\, I_C)$ } 
        }}

    \thicklines
    \put(305,240){\vector(0,1){40}}
    \put(285,290){\parbox[b]{60pt}
      {\small conditioned\\ per-event\\ statistics}}

    \put(305,200){\vector(0,-1){40}}
    \put(285,120){\parbox[b]{60pt}
      {\small conditioned\\ sample\\ hierarchy}}

  \end{picture}
\end{center}
\vspace*{15pt}
\caption{Example of a simple univariate sampling hierarchy, Version 1.1}
\end{figure}
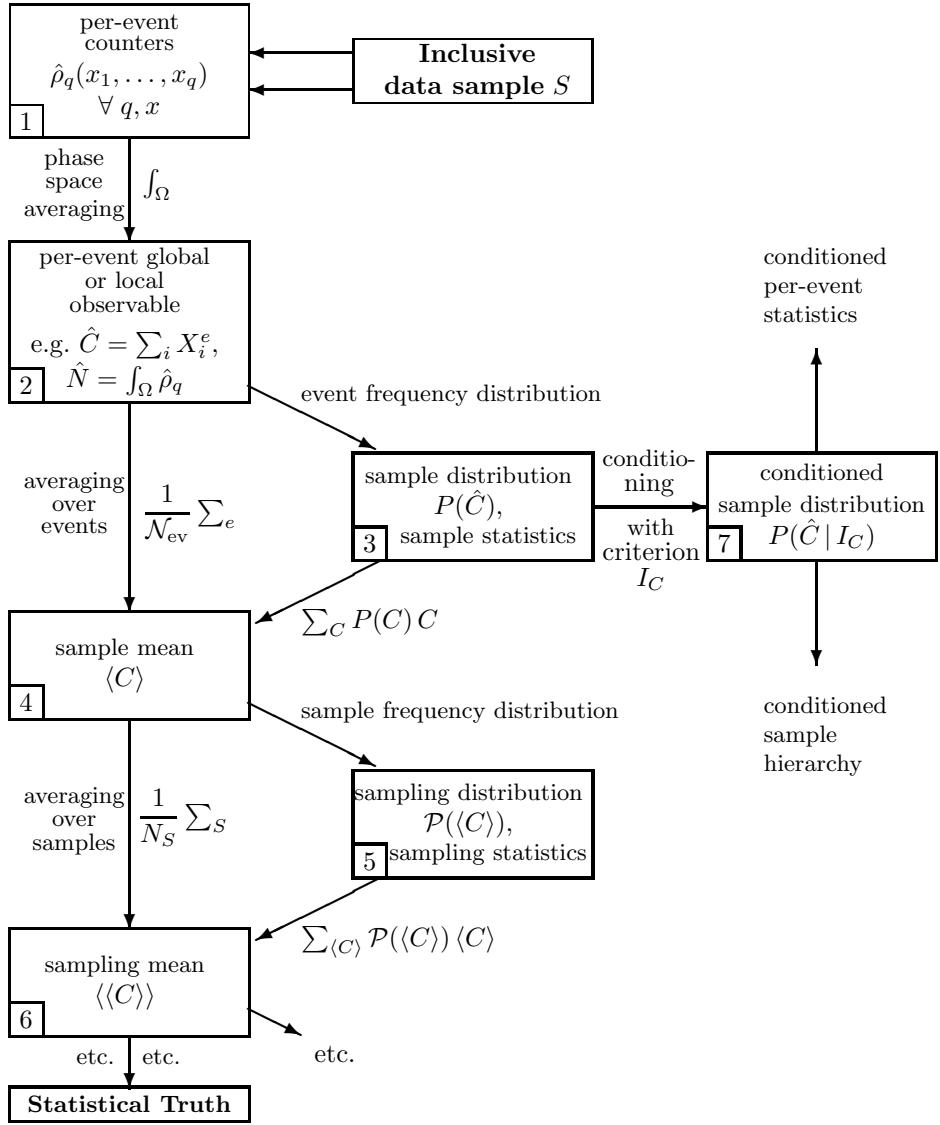


\end{document}